\documentclass[twocolumn,aps,showpacs,nofootinbib,floatfix]{revtex4}
\usepackage{graphicx}
\usepackage{amsmath,amssymb}
\usepackage{footmisc}

\begin{document}

\title{The Astrophysical Uncertainties Of Dark Matter Direct Detection Experiments}

\author{Christopher McCabe}
\email{mccabe@thphys.ox.ac.uk}
\affiliation{Rudolf Peierls Centre for Theoretical Physics, University of Oxford,\\1 Keble Road, Oxford, OX1 3NP, UK}

\vskip 0.5cm

\date{\today}

\begin{abstract}
The effects of astrophysical uncertainties on the exclusion limits at dark matter direct detection experiments are investigated for three scenarios: elastic, momentum dependent and inelastically scattering dark matter. We find that varying the dark matter galactic escape velocity and the Sun's circular velocity can lead to significant variations in the exclusion limits for light ($\lesssim10$ GeV) elastic and inelastic scattering dark matter. We also calculate the limits using one hundred velocity distributions extracted from the Via Lactea II and GHALO N-body simulations and find that a Maxwell-Boltzmann distribution with the same astrophysical parameters generally sets less constraining limits. The elastic and momentum dependent limits remain robust for masses $\gtrsim50$ GeV under variations of the astrophysical parameters and the form of the velocity distribution.
\end{abstract}

\pacs{95.35.+d  \hspace{7.2cm} Preprint: OUTP-10-09-P}

\maketitle


\section{Introduction}
Identifying the dark matter in our Universe is one of the great challenges facing both theorists and experimentalists. Dark matter direct detection experiments are now probing dark matter-nucleon cross sections expected of Weakly Interacting Massive Particles (WIMPs) and the results from the CDMS II \cite{Ahmed:2009zw} and CoGeNT \cite{Aalseth:2010vx} experiments may be hinting that a conclusive detection of dark matter is within reach of the next generation of experiments. The DAMA collaboration \cite{Bernabei:2010mq} have also continued to measure an annual modulation signal, observed over the past thirteen years with an $8.9 \sigma$ significance which has many of the properties expected from a dark matter origin. Although now inconsistent with the null results from other experiments assuming a vanilla elastically scattering WIMP, this persistent signal has inspired many models \cite{TuckerSmith:2001hy, Bai:2009cd, Feldstein:2009tr, Chang:2009yt, Frandsen:2009mi, Masso:2009mu, Kopp:2009et,Essig:2010ye,Graham:2010ca} which can give rise to dramatically different experimental signals from the usual assumptions of elastically scattering spin-independent or spin-dependent interactions. Although these models may be capable of reconciling the DAMA result with other experiments, we consider them to be interesting, independent of the DAMA result, since they serve to highlight the different possible signals that could be produced at dark matter direct detection experiments.

As well as depending on particle physics models, the recoil spectrum and scattering rate depend on the astrophysical parameters and the form of the local velocity distribution the particle dark matter in the galactic halo takes. Unfortunately, these parameters still suffer from large uncertainties, while high resolution numerical N-body simulations have shown that the velocity distribution deviates significantly from the default choice, namely, a Maxwell-Boltzmann distribution.

Given this, in this paper we investigate the effect of astrophysical uncertainties on three scenarios: elastic spin-independent scattering; and, as proxies for the many varied models, inelastic and momentum dependent (aka form factor) spin-independent scattering. We vary the galactic escape velocity, the Sun's velocity relative to the galactic centre, and the local dark matter density within their experimental errors. We then compare the exclusion limits for these scenarios assuming a Maxwell-Boltzmann velocity distribution with those extracted from two of the highest resolution dark matter N-body simulations, Via Lactea II and GHALO. We particularly focus on four experiments with different target nuclei which span a large range of atomic number $A$ to compare the relative change at different experiments: CDMS II-Si; CDMS II-Ge; XENON10; and CRESST-II, with silicon ($A=28$), germanium ($A=72$), xenon ($A=131$) and tungsten ($A=184$) target nuclei respectively.\footnote{Since we are seeking to understand the effects of the astrophysical uncertainties, we do not specifically address whether astrophysical uncertainties can improve the fit between the CoGeNT and DAMA signals. For analyses in this direction, see \cite{Fitzpatrick:2010em,Chang:2010yk}.}

\begin{figure}[t]
\centering
\includegraphics[height=2.0in]{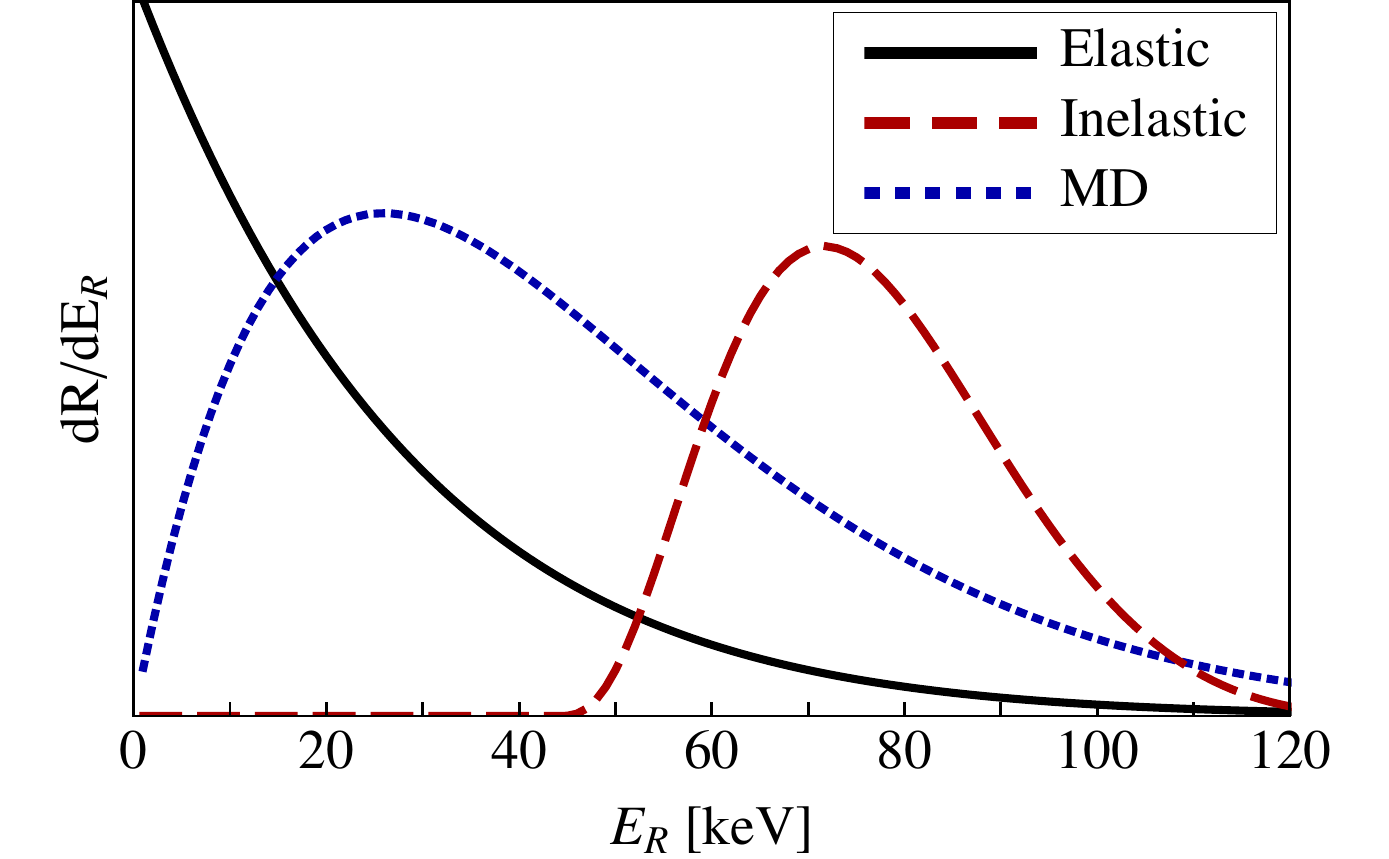}\caption{(Colour online). The differential scattering rate at a germanium detector for $M_{\chi}=100$ GeV for elastic, inelastic and momentum dependent (MD) spin-independent scattering. The normalisation of each curve is different.}
\label{3dRdE}
\end{figure}

In Fig.\ \ref{3dRdE} we have plotted the differential scattering rate at a germanium detector to illustrate the differences between the three scenarios. Inelastic scattering of the dark matter to a state $\sim100$ keV heavier \cite{TuckerSmith:2001hy} changes the phenomenology in three principal ways: it leads to a suppression of scattering rates at low mass target nuclei experiments; an increase in the ratio of the modulated to unmodulated scattering rate; and a kinematical suppression of the recoil spectrum at low recoil energies. Although often invoked as a way of reconciling the modulation signal seen by DAMA with the null results at other experiments \cite{TuckerSmith:2004jv, Chang:2008gd, MarchRussell:2008dy, Cline:2009xd, SchmidtHoberg:2009gn}, the latest (preliminary) results from CRESST-II \cite{CRESSTWONDER} tentatively rule out this interpretation. However, having this splitting is natural in many models \cite{Cui:2009xq,Alves:2009nf,Arina:2009um,ArkaniHamed:2000bq,MarchRussell:2004uf,Thomas:2007bu,MarchRussell:2009aq,ArkaniHamed:2008qn,Cholis:2009va,Gherghetta:2010cq} and leads to much other interesting phenomenology, for example through capture in astrophysical bodies \cite{Nussinov:2009ft, Menon:2009qj, Shu:2010ta,McCullough:2010ai, Hooper:2010es}. Therefore, we emphasise that even if inelastic scattering can not reconcile the DAMA result with other experiments, it remains an interesting and viable possibility to explore.

In momentum dependent scattering \cite{Feldstein:2009tr, Chang:2009yt}, the differential cross section is suppressed by additional powers of $q^2=2M_NE_R$, leading to a suppression of the exclusion limits from experiments which rely on having a sensitivity at low recoil energies, in our case XENON10 and CDMS II-Si. This dependence can arise for example through dipole or charge radius interactions \cite{An:2010kc}.

In Section \ref{Review} we briefly review the theory required to calculate scattering rates, then in Section \ref{astrouncert} we review our current knowledge of the relevant astrophysical parameters and demonstrate and explain their effect on the experimental exclusion curves for elastic, inelastic and momentum dependent scattering. In Section \ref{VLGHALO} we examine the effect of using velocity distributions from the Via Lactea II and GHALO simulations. Finally, Appendix \ref{experiments} summarises the relevant experimental details, where in particular, we call attention to the reanalysed early CDMS II runs and discuss uncertainties affecting the XENON10 exclusion limits at low masses.

\section{Review of standard formalism}\label{Review}
In this section we briefly review the theory needed to calculate the number of events expected at a given experiment. The differential scattering rate is given by
\begin{equation}\label{dRdE1}
\frac{dR}{dE_R}=N_T\frac{\rho_{\chi}}{M_{\chi}}\int^\infty_{v_{min}} v f_{\oplus}(\vec{v},\vec{v}_e) \frac{d \sigma}{d E_R}d^3v ,
\end{equation}
where $N_T$ is the number of target nuclei, $\rho_{\chi}$ is the local dark matter density, $M_{\chi}$ is the dark matter mass, $v=\lvert \vec{v}\rvert$, and $f_{\oplus}(\vec{v},\vec{v}_e)$ is the local dark matter velocity distribution in the detector rest frame. For a nucleus to recoil with an energy $E_R$, the incident dark matter particle must have a minimum speed given by:
\begin{equation}\label{vmin}
\frac{v_{min}}{c}=\sqrt{\frac{1}{2 M_N E_R}}\left(\frac{M_NE_R}{\mu_N}+\delta\right).
\end{equation}
Here $M_N$ is the nucleus mass, $\mu_N$ the dark matter-nucleus reduced mass, and we include $\delta$, the mass difference between the incoming and outgoing dark matter particle, to allow for inelastic scattering.

We will concentrate on coherent spin-independent interactions in the remainder of this paper, in which case, the dark matter-nucleus differential cross section is given by
\begin{equation}\label{dsigmadE}
\frac{d \sigma}{d E_R}=\frac{1}{2 v^2}\frac{M_N \sigma_n}{\mu^2_{ne}}{\frac{(f_p Z + f_n (A-Z))^2}{{f_n}^2}} F^2(E_R),
\end{equation}
where $\sigma_n$ is the dark matter-neutron cross section at zero momentum transfer in the elastic limit, $\mu_{ne}$ is the dark matter-nucleon reduced mass and $f_n$ and $f_p$ parameterize the relative dark matter coupling to neutrons and protons respectively. For simplicity, we shall assume $f_n=f_p$ for the remainder of this paper, but one should bear in mind that in specific particle physics models, this may not be the case. Fortunately the rescaling of the cross section for different values of $f_n$ and $f_p$ is trivial. $F(E_R)$ is the nucleus form factor which for tungsten and xenon nuclei, we take as the Fermi Two-Parameter distribution with parameters taken from the Appendix of \cite{Duda:2006uk}. For germanium and silicon we use the Lewin-Smith \cite{Lewin:1995rx} parameterisation of the Helm form factor: 
\begin{equation}
F^2(E_R)=\left(\frac{3 j_1(qR)}{qR}\right)^2 e^{-q^2 s^2}
\end{equation}
where $q=\sqrt{2 M_N E_R}$, $R=\sqrt{c^2+\frac{7}{3} \pi^2 a^2 -5 s^2}$, $c=1.23 A^{1/3}-0.60$ fm, $s=0.9$ fm and $a=0.52$ fm. We find that this parameterisation agrees well (to within 2\% or better) with the `spherical Bessel expansion' and `sum of Gaussian expansion' from \cite{Duda:2006uk}, in stark contrast to the commonly used values $R=\sqrt{r^2-5 s^2}$, $s=1$ fm and $r=1.2 A^{1/3}$ which overestimates $F^2(E_R)$ by up to $20\%$ at $E_R=100$ keV. As pointed out in \cite{Duda:2006uk}, the uncertainty in the exclusion limits due to the form factor parameterisation can be significantly reduced by a judicious choice, leaving, as we will show, uncertainties in the astrophysical parameters as the dominant source of uncertainty in the exclusion limits.

Finally, to allow for momentum dependent scattering, we follow \cite{Chang:2009yt} by parameterising the momentum dependence by replacing the differential cross section with
\begin{equation}\label{FFdsigmadE}
\frac{d \sigma_{}}{d E_R}\rightarrow\left(\frac{q^2}{q^2_{ref}}\right)^n\frac{d \sigma}{d E_R},
\end{equation}
where $q_{ref}=100$ MeV. Here we only consider the case $n=1$, although models with $n=2$ have also been constructed.

\subsection{Astrophysical formalism}\label{astroform}

The simplest model of the dark matter distribution is the so called `standard halo model' (SHM), which assumes an isotropic, isothermal sphere for the dark matter distribution, and leads to a Maxwell-Boltzmann velocity distribution  in the galactic frame, truncated at the galaxy escape velocity $v_{esc}$:
\begin{equation}\label{fgal}
f(\vec{v}) = \left\{
\begin{array}{l@{\qquad}l}
\frac{1}{N} \left(e^{-v^2 / v_0^2} - e^{-v_{esc}^2 / v_0^2}\right)
& v < v_{esc} \\
0 & v > v_{esc}
\end{array}\right.
\end{equation}
Here $N$ is a normalisation constant and $v_0$ is the circular speed of the Sun about the centre of the galaxy (which throughout this paper we assume is the same as the local standard of rest). High resolution N-body simulations of the galactic dark matter structure have shown that the dark matter velocity distribution departs from the SHM \cite{Hansen:2005yj,Kuhlen:2009vh, Vogelsberger:2008qb}, and the effects can be large, especially in scenarios such as inelastic dark matter where only the high velocity tail of the velocity distribution is sampled \cite{Green:2000jg,Green:2002ht,Green:2003yh,Belli:2002yt,Bottino:2005qj,Vergados:2007nc,Fairbairn:2008gz, MarchRussell:2008dy,Ling:2009eh}. Therefore in Section \ref{VLGHALO} we explore how the exclusion limits vary when we use data from the Via Lactea II \cite{Diemand:2008in} and GHALO \cite{Stadel:2008pn} simulations, two of the highest resolution simulations of galactic dark matter structure.

The velocity distribution entering Eqn.\ \ref{dRdE1} is evaluated in the frame of the detector, therefore we need to transform the galactic velocity distibution (Eqn. \ref{fgal}) to the Earth frame.  $f_{\oplus}(\vec{v},\vec{v}_e)$ and $f(\vec{v})$ are simply related through:
\begin{equation}\label{fearthgal}
 f_{\oplus}(\vec{v},\vec{v}_e) = f(\vec{v}+\vec{v}_e)
 \end{equation}
where $\vec{v}_e=\vec{v}_ \odot +\vec{v}_\oplus$. $\vec{v}_\odot=\vec{v}_0+\vec{v}_\circledast$ is the sum of the Sun's circular velocity $\vec{v}_0$ relative to the galactic centre, and peculiar velocity $\vec{v}_\circledast$ relative to Sun's circular velocity. $\vec{v}_\oplus$ is the Earth's velocity relative to the Sun's rest frame which we calculate following the prescription in Appendix B of Ref. \cite{Lewin:1995rx}.  Combining Eqns.\ \ref{dRdE1}, \ref{dsigmadE} and \ref{fearthgal}, the relevant quantity which enters into the integration in $dR/dE_R$ is:
\begin{equation}\label{zetaER}
\zeta(E_R)=\int^\infty_{v_{min}} \frac{d^3v}{v} f(\vec{v}+\vec{v}_e).
\end{equation}
In Appendix \ref{analytic_integration} we present an analytic expression for $\zeta(E_R)$ for the SHM velocity distribution, as defined by Eqn. \ref{fgal}, which to the best of our knowledge has not previously been presented in the literature.


\section{Astrophysical Uncertainties}\label{astrouncert}

\begin{figure*}[t]
\centering
\includegraphics[height=6.3in]{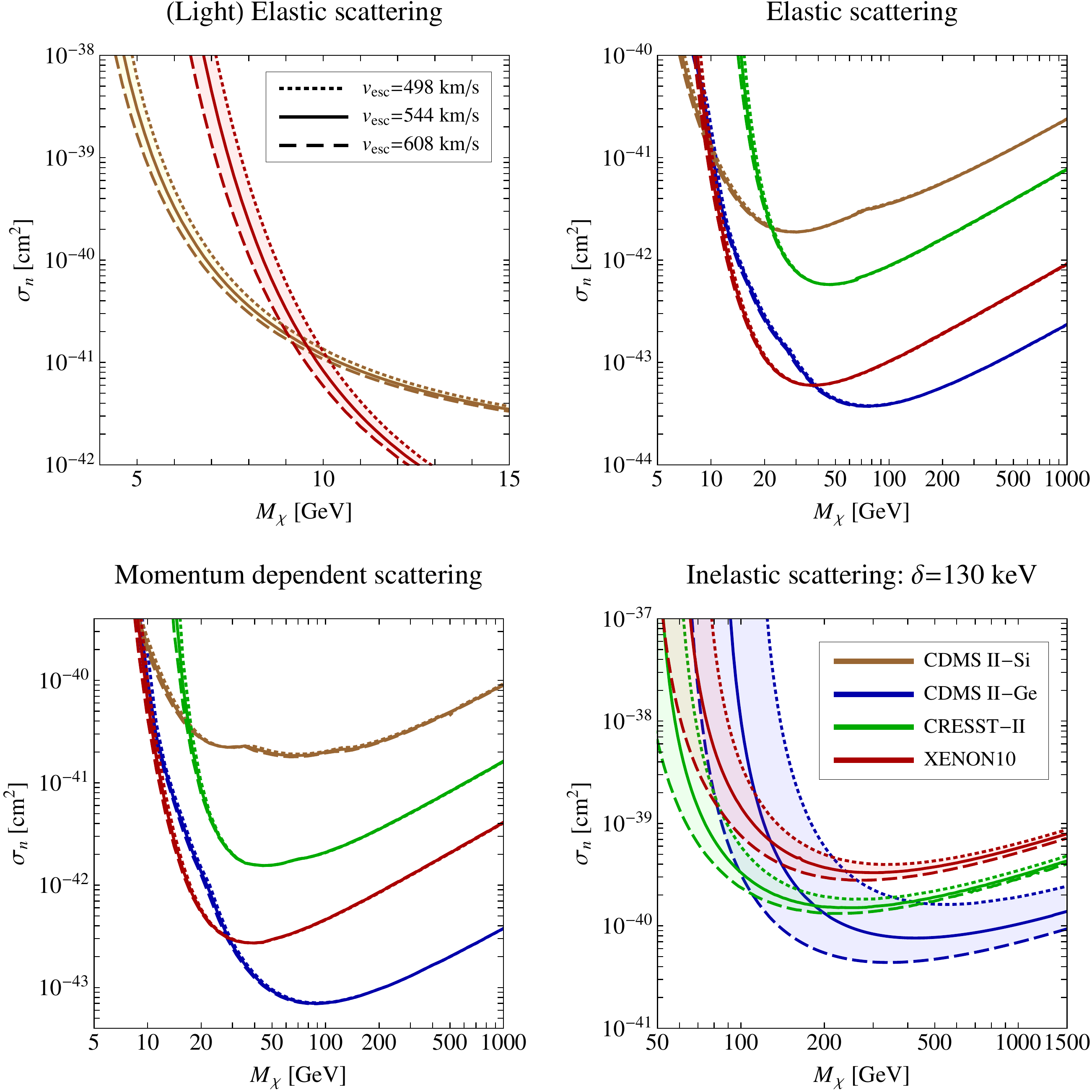}\caption{(Colour online). Varying $v_{esc}$ within the $90\%$ confidence limit found by the RAVE survey: $v_{esc}=498$ km/s (dotted), $544$ km/s (solid) and $608$ km/s (dashed). Varying $v_{esc}$ has a negligible effect on the limits for momentum dependent and elastically scattering dark matter except at low masses, where the limits are shifted horizontally by $\sim0.5$ GeV. For inelastically scattering dark matter, the effect is much larger, especially for CDMS II-Ge. The two principal features to note are a shift left (right) and down (up) when $v_{esc}$ is increased (decreased).}
\label{v_esc}
\end{figure*}

The quantities entering into this calculation from astrophysics are: $\rho_{\chi}$, the local dark matter density, with canonical value $\rho_{\chi}=0.3$ GeV/cm$^3$;  the Sun's circular velocity with respect to the galactic rest frame with fiducial value $\vec{v}_0=(0, 220 ,0)$ km/s; the Sun's peculiar velocity relative to the Sun's circular velocity, which is usually given by $\vec{v}_ \circledast =(10.0, 5.25 ,7.17)$ km/s \cite{Dehnen:1997cq}; and the escape velocity $v_{esc}$, which is often chosen in the range from $500$ km/s to $650$ km/s. In this section we will discuss recent determinations of these parameters and investigate the changes in the exclusion limits when we vary them within their allowed ranges, for elastic, momentum dependent and inelastic scattering dark matter with $\delta=130$ keV.

\subsection{The galactic escape velocity}\label{sectionvesc}

The RAVE survey \cite{Smith:2006ym} has determined a $90\%$ confidence interval for the galactic escape velocity of $498\text{ km/s}< v_{esc} <608\text{ km/s}$ with a median likelihood of $544$ km/s. In Fig. \ref{v_esc} we plot the exclusion limits for CDMS II-Si in brown, CDMS II-Ge in blue, CRESST-II in green and XENON10 in red for elastically scattering (left and right upper panels), momentum dependent (lower left panel) and inelastically scattering dark matter with a mass splitting $\delta=130$ keV (lower right panel) for three values of $v_{esc}$; $498$ km/s (dotted); $544$ km/s (solid); and $608$ km/s (dashed) while keeping $\rho_{\chi}=0.3$ GeV/cm$^3$ and $v_0=220$ km/s fixed. For clarity, we have only shown the limits for CDMS II-Si and XENON10 at low masses (upper left panel), since the CDMS II-Ge limits behave similarly to those from XENON10 in this mass range. CDMS II-Si is not able to set any limits for inelastic dark matter with $\delta=130$ keV since its low mass means the minimum speed required to scatter off a nucleus is higher than the galactic escape velocity.

The limits for momentum dependent and elastically scattering dark matter are very insensitive to $v_{esc}$ for all experiments and it is only for masses $M_{\chi}\lesssim10$ GeV that the three lines can be separately resolved. In comparison, inelastically scattering dark matter is much more sensitive, with two principal features to note: the curves shift left (right) and down (up) when $v_{esc}$ is increased (decreased).

\begin{figure}[t]
\centering
\includegraphics[height=2.0in]{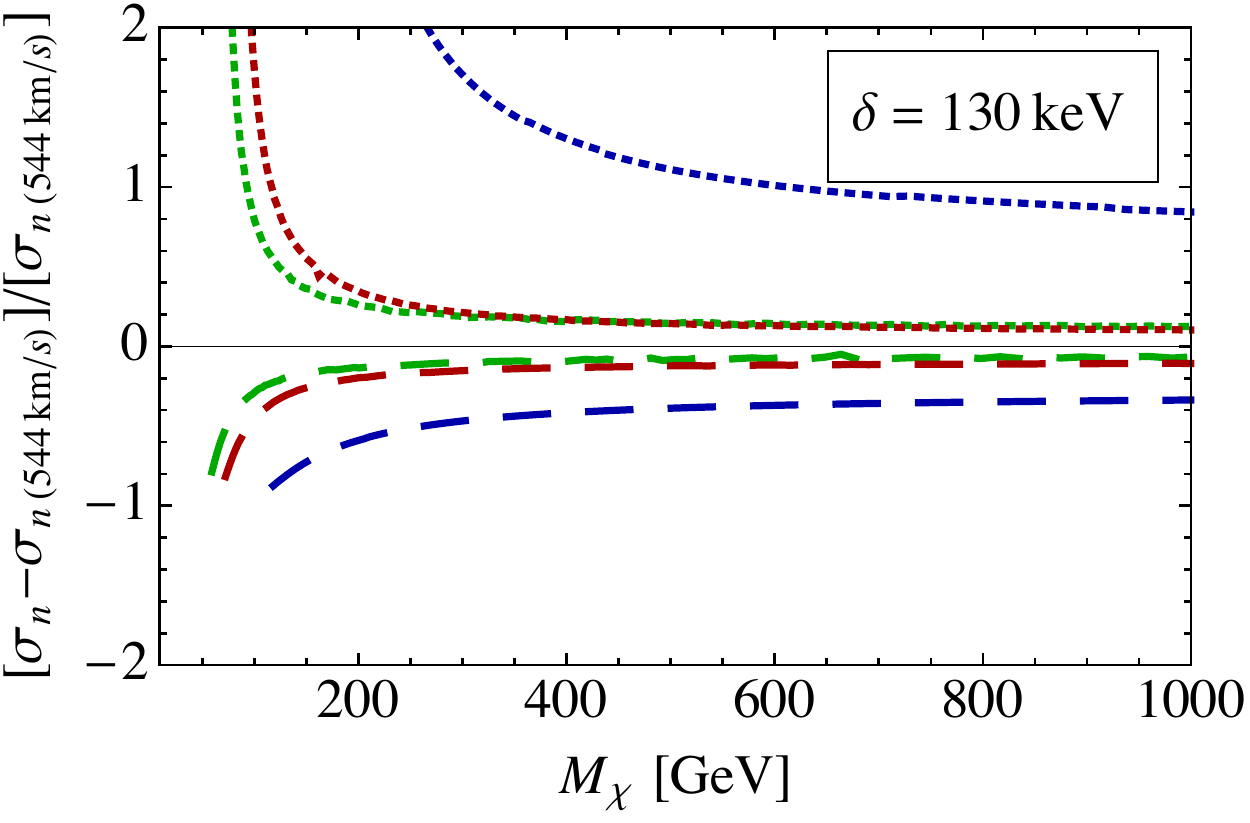}\caption{(Colour online). The fractional change in $\sigma_n$ relative to value at $v_{esc}=544$ km/s for $v_{esc}=498$ km/s (dotted) and $v_{esc}=608$ km/s (dashed) for inelastically scattering dark matter with $\delta=130$ keV. The blue, red and green lines show the change for CDMS II-Ge, XENON10 and CRESST-II respectively.}
\label{ratiovesc}
\end{figure}

In Fig. \ref{ratiovesc} the fractional change in $\sigma_n$ relative to the value at $v_{esc}=544$ km/s is plotted as a function of mass for the inelastically scattering case. The dotted and dashed lines show the ratio for $v_{esc}=498$ km/s and $608$ km/s respectively. We see that the XENON10 and CRESST-II limits change in a similar way, varying by $<20\%$ at high masses,  and remaining fairly constant until $\sim100$ GeV, when a very rapid increase occurs. In comparison, decreasing the escape speed dramatically weakens the CDMS II-Ge limits, by $80\%$ at high masses ($M_{\chi}\sim1000$ GeV) and larger at lower masses. Increasing $v_{esc}$ has a less dramatic effect with a change of $\sim40\%$ for CDMS II-Ge until $\sim150$ GeV when a rapid increase happens. These effects are simple to understand qualitatively and quantitively.
 
We will first explain why the curves move vertically in Fig. \ref{v_esc}. If there is a lower escape velocity, there will be fewer particles in the halo capable of scattering, therefore a larger $\sigma_n$ is required to produce the same number of events at a given experiment. Since the Maxwell-Boltzmann distribution is exponentially suppressed at high velocities, decreasing $v_{esc}$ will have a very small effect in the number of particles able to scatter, unless $v_{min}$ is close to $v_{esc}$, in which case the fractional change can be substantial. For elastic and momentum dependent scattering dark matter, this occurs at $M_{\chi} \lesssim 10$ GeV, and is true for all masses when $\delta=130$ keV. CDMS II-Ge is particularly affected relative to XENON10 and CRESST-II because its germanium target is much lighter than xenon and tungsten, so $v_{min}$ is higher, and much closer to $v_{esc}$. Similarly if $v_{esc}$ is raised, there are extra particles in the halo which can scatter, so a lower $\sigma_n$ will produce the same number of events in an experiment. Due to the exponential suppression of the Maxwell-Boltzmann distribution, increasing $v_{esc}$ has a smaller affect than decreasing it by a similar amount.

Unfortunately, the intricacies of the `pmax' method \cite{yellin} for setting exclusion limits prevents very accurate estimates. However, we can give a quantitive estimate which should serve as a useful guide to the size of shift expected when $v_{esc}$ is varied. Since the combination $\sigma_n \zeta$ enters $dR/dE_R$ (from Eqns.\ \ref{dsigmadE} and \ref{zetaER}), the vertical shift is given by
\begin{equation}
\Delta \sigma^i_j\equiv\frac{\sigma_{n (i)}-\sigma_{n (j)}}{\sigma_{n (j)}}=\frac{\zeta_{(j)}-\zeta_{(i)}}{\zeta_{(i)}}
\end{equation}
 where an analytic expression for $\zeta$ is given by Eqn.\ \ref{analytic2}. 
 
For example, for CRESST-II with $v_{min}$ from Eqn.\ \ref{vmin}, $M_{\chi}=500$ GeV, $\delta=130$ keV, and $E_R=25$ keV,\footnote{The peak energy of the recoil spectrum $dR/dE_R$ \cite{Chang:2008gd}\label{first_footnote}} we find $\Delta \sigma^{498km/s}_{544km/s}=0.13$ and $\Delta \sigma^{608km/s}_{544km/s}=-0.08$, in good agreement with the values $0.15$ and  $-0.08$ presented in Fig. \ref{ratiovesc}. In comparison for CDMS II-Ge, with $M_{\chi}=500$ GeV, $\delta=130$ keV and $E_R=80$ keV \footnotemark[\value{footnote}], we find $\Delta \sigma^{498km/s}_{544km/s}=0.75$ and $\Delta \sigma^{608km/s}_{544km/s}=-0.29$. This compares to the values of $1.1$ and $-0.39$ shown in Fig. \ref{ratiovesc}, a difference of around $40\%$.  

To understand the horizontal shift, we will consider the energy of the recoiling nucleus, which is ultimately what experiments measure. In the Earth frame, the nuclear recoil energy is given by 
\begin{equation}\label{Erecoil}
E_R=\frac{2 \mu_N^2 v_{\chi}^2 \cos^2\theta_R}{M_N}
\end{equation}
where $v_{\chi}$ is the dark matter speed in the Earth's frame and $\theta_R$ is the nuclear recoil angle. If the recoil energy is the same, we expect a similar signal at an experiment. Increasing $v_{esc}$ will increase the average speed of dark matter particles in the halo, but this increase will only be perceptible if $v_{min}$ is close to $v_{esc}$, as for inelastic and light dark matter. When the average speed increases, the average kinetic energy will be the same for a lighter dark matter particle, and since we expect the nuclear recoil energy to depend on a simple way on the dark matter's initial kinetic energy, the exclusion curves shift to the left.\footnote{A similar argument is presented in \cite{Green:2008rd, Green:2010ri}}

 \begin{figure*}[t]
\centering
\includegraphics[height=6.3in]{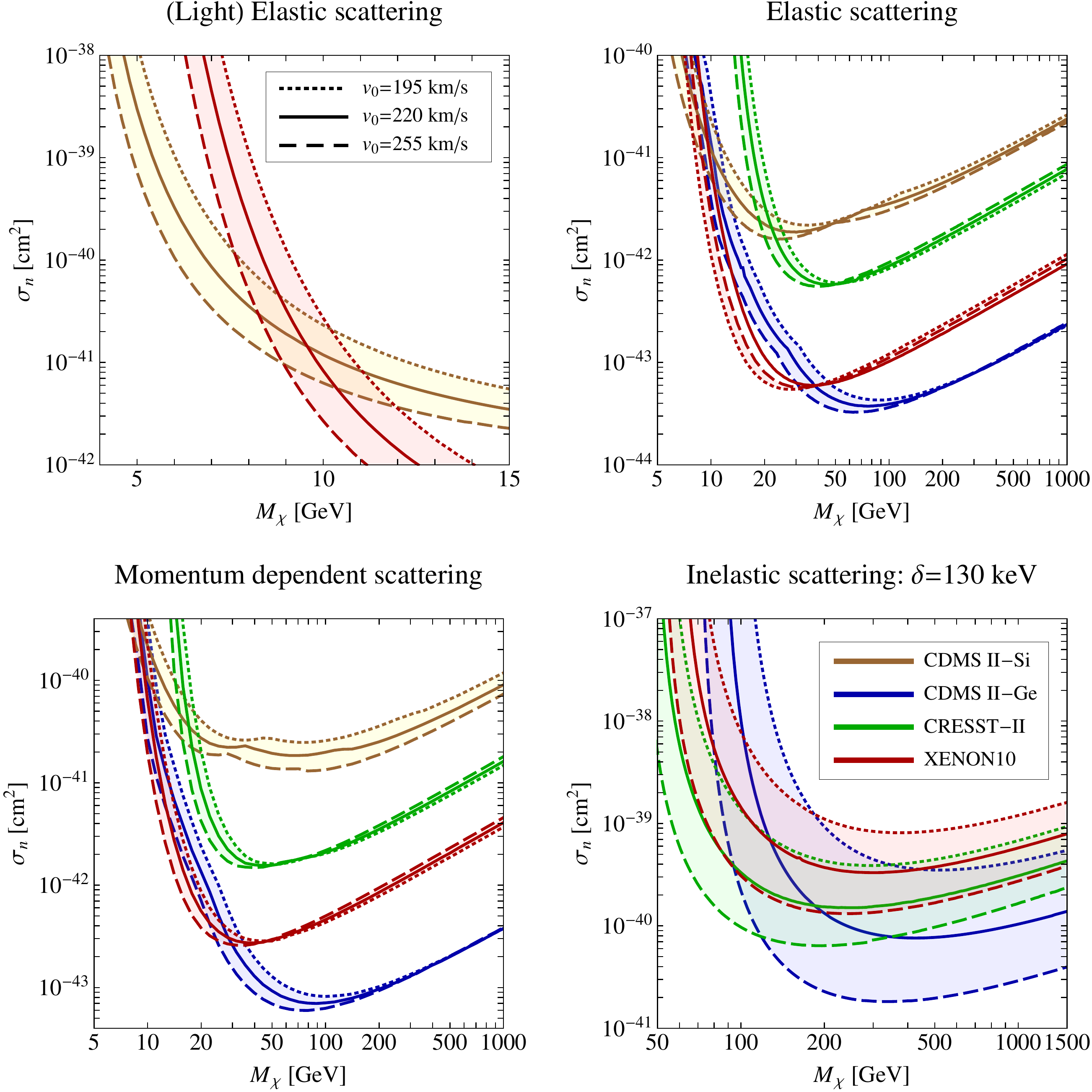}\caption{(Colour online). The exclusion limits for $v_0=195$ km/s (dotted), $v_0=220$ km/s (solid) and $v_0=255$ km/s (dashed). Varying $v_0$ leads to a larger vertical shift in the exclusion limits compared to varying $v_{esc}$ for all panels (cf.\ Fig.\ \ref{v_esc}). The horizontal shift for all experiments at masses $\sim10$ GeV for elastic and momentum dependent scattering is $\sim1-2$ GeV, larger than when $v_{esc}$ is varied (cf.\ Fig.\ \ref{v_esc}). In contrast, when varying $v_0$ in the inelastic case, the horizontal shift is smaller compared to varying $v_{esc}$.
}
\label{v0}
\end{figure*}
 
Quantitively, the mass shift is found from differentiating Eqn.\ \ref{Erecoil}, leading to 
\begin{equation}\label{deltaM}
\frac{\Delta M_{\chi}}{M_{\chi}}=-(1+\frac{M_{\chi}}{M_N})\frac{\Delta v_{\chi}}{v_{\chi}}.
\end{equation}
where $\Delta M_{\chi}=M_{\chi 2}-M_{\chi 1}$, and $\Delta v_{\chi}=v_{\chi 2}-v_{\chi 1}$ corresponds to the shift in the average speed of the dark matter responsible for measurable events.

For CDMS II-Ge with $\delta=130$ keV, $M_{\chi}=90$ GeV, the only particles which can scatter have a speed $v_{\chi}\sim v_{esc}+v_e=800$ km/s. Changing $v_{esc}$ from $544$ km/s by $\pm64$ km/s leads to $\Delta v_{\chi}=\pm64$ km/s, which from Eqn.\ \ref{deltaM} gives $\Delta M_{\chi}\sim\mp17$ GeV, in reasonable agreement with what we find in Fig.\ \ref{v_esc}. For the elastic case $\Delta v_{\chi}\sim0$ because the average speed doesn't depend sensitively on the escape speed, unless $v_{min}\sim v_{esc}+v_e$ which occurs at $M_{\chi}\sim10$ GeV. For XENON10 at $M_{\chi}=7$ GeV, $\delta=0$ keV, $v_{\chi}=800$ km/s and $\Delta v_{\chi}=\pm64$ km/s, this leads to $\Delta M_{\chi}\sim\mp0.6$ GeV, again in reasonable agreement with Fig.\ \ref{v_esc}.

\subsection{The Sun's peculiar velocity}

The peculiar velocity $\vec{v}_\circledast=(U,V,W)_\circledast$ is often assumed to be well known: $\vec{v}_ \circledast =(10.0, 5.25 ,7.17)$ km/s with small errors $\sim 0.5$ km/s \cite{Dehnen:1997cq}. However three recent papers have suggested that $V_ \circledast$ may have been underestimated by $\sim 7$ km/s \cite{McMillan:2009yr, Binney:2009ym, Schoenrich:2009bx}. Ref.\cite{Schoenrich:2009bx} found that the classical determination in \cite{Dehnen:1997cq} underestimated $V_ \circledast$ by ignoring the metallicity gradient in the Milky Way disc. Furthermore when  $\vec{v}_\circledast$ is increased, \cite{McMillan:2009yr} and \cite{ Binney:2009ym} found better fits to the observations of masers in high mass star forming regions and the Milky Way distribution functions respectively. Therefore unless otherwise stated, in the remainder of this paper we will use this newly determined value for the Sun's peculiar velocity: $\vec{v}_ \circledast=(11.1, 12.24 ,7.25)$ km/s with uncertainties  $\sim(1, 2 ,0.5)$ km/s.

\subsection{The Sun's circular velocity}\label{Suncirc}

\begin{figure}[t]
\centering
\includegraphics[width=3.3in]{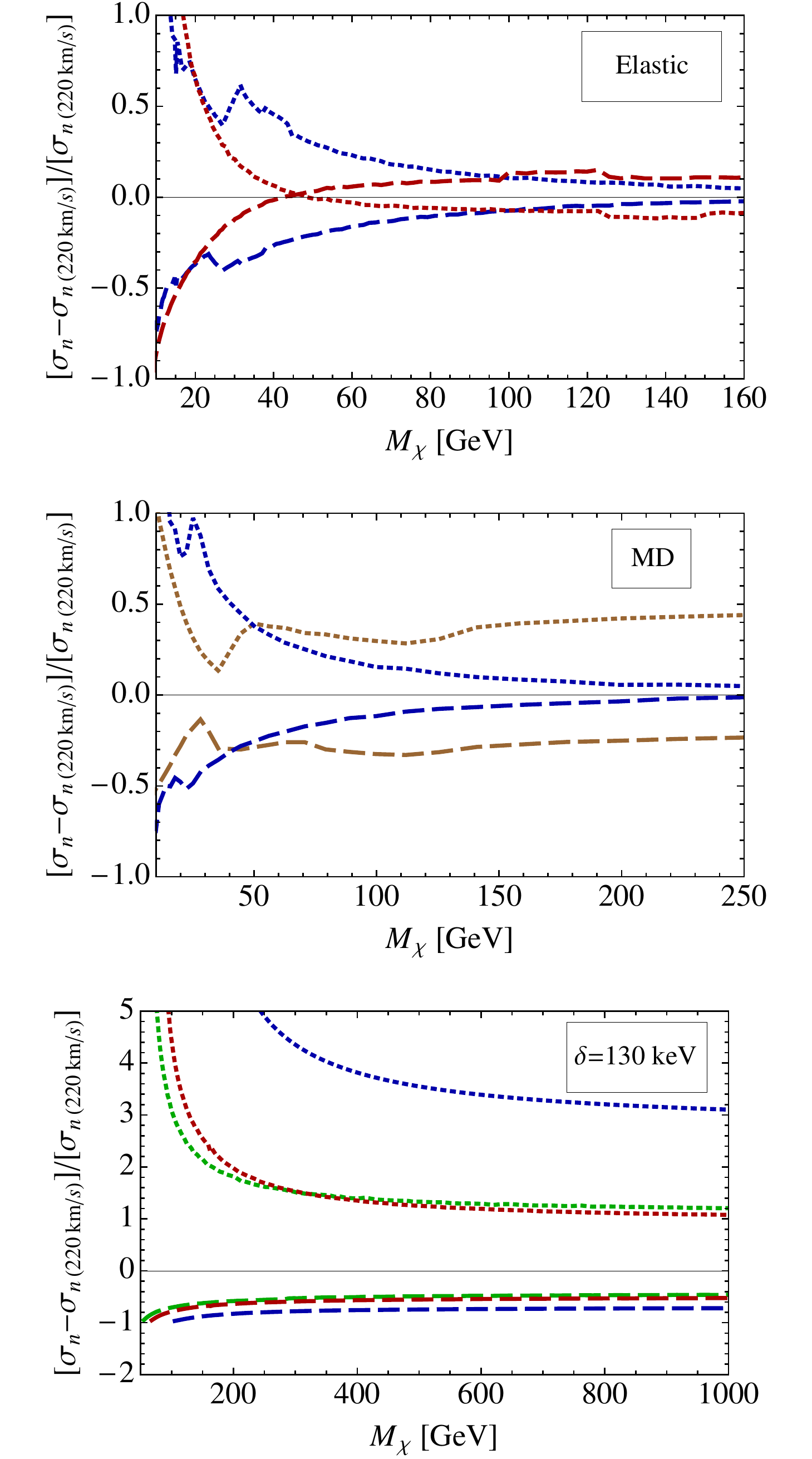}\caption{(Colour online). The fractional change in $\sigma_n$ relative to value at $v_{0}=220$ km/s for $v_{0}=195$ km/s (dotted) and $v_{0}=255$ km/s (dashed) for elastically (top panel), momentum dependent (middle panel) and inelastically scattering dark matter with $\delta=130$ keV (bottom panel). The blue, brown, red and green lines show the change for CDMS II-Ge, CDMS II-Si, XENON10 and CRESST-II respectively. For clarity we only show the change for CDMS II-Ge and XENON10 in the top panel and CDMS II-Ge and CDMS II-Si in the middle panel.}
\label{v0fraction}
\end{figure}

The Sun's circular velocity can be determined in a number of ways, which give slightly different values from the fiducial value of $v_0=220$ km/s. The ratio of the total velocity of the Sun about the galactic centre to the distance of the Sun from the galactic centre has been determined from the apparent proper motion of Sgr A$^*$ to be $v_{\odot}/R_0=(v_{0}+V_{\circledast})/R_0=30.2 \pm 0.2$ km s$^{-1}$kpc$^{-1}$ \cite{Reid:2004rd}. Unfortunately $R_0$ is still poorly determined with the most recent studies of stellar orbits obtaining $R_0=8.4\pm0.4$ kpc \cite{Ghez:2008ms} and $R_0=8.33\pm0.35$ kpc \cite{Gillessen:2008qv}. Combining these results, and using the new value of $V_{\circledast}=12.24$ km/s, we find $v_0=242\pm12$ km/s and $v_0=239\pm11$ km/s respectively. This is in comparison to the analysis of the GD-1 stellar stream which found $v_0=221\pm18$ km/s \cite{Koposov:2009hn}, and the best fit to masers in the high mass forming regions which found the range $v_0=225\pm29$ km/s \cite{McMillan:2009yr}. Wary of the possibility of unknown systematic errors affecting one of these measurements of $v_0$, we take a conservative approach by giving each the same weight and hence in Fig.\ \ref{v0} we investigate how the exclusion limits change for three values of $v_0$; $195$ km/s (dotted); $220$ km/s (solid); and $255$ km/s (dashed), while keeping $\rho_{\chi}=0.3$ GeV/cm$^3$ and $v_{esc}=544$ km/s fixed.  Again the left and right upper panels shows elastically scattering dark matter while the lower left and right panels show momentum dependent and inelastically scattering dark matter for $\delta=130$ keV. We see that varying $v_0$ has more of an effect on $\sigma_n$ than changing $v_{esc}$ for all cases (cf.\ Fig.\ \ref{v_esc}), and that once again there are two principal features to note: increasing (decreasing) $v_0$ causes the exclusion curves to shift down (up) and left (right).

Changing $v_0$ has two effects: $v_0$ is needed to boost from the galactic frame to the Earth frame (Eqn.\ \ref{fearthgal}), so increasing $v_0$ means there are more particles at speeds above $v_{min}$ available to scatter in the Earth's frame; and in the SHM, $v_0$ determines the dark matter dispersion velocity ($\sigma_{\text{dis}}\equiv\sqrt{\frac{3}{2}}v_0$), so increasing $v_0$ also increases the range of the speed of particles in the halo, with more at higher and lower speeds. Therefore if $v_{min}$ is close to $v_{esc}$, as for inelastic dark matter and light dark matter, changing $v_0$ has a similar effect to changing $v_{esc}$ since it also increases or decreases the number of particles which can scatter. However since we are not just changing the tail of an exponential distribution, but rather its overall shape and position, the effect of changing $v_0$ is larger than varying $v_{esc}$. This is demonstrated in the bottom panel of Fig.\ \ref{v0fraction} where we have plotted the fractional change in $\sigma_n$ relative to the value at $v_0=220$ km/s for $v_0=195$ km/s (dotted) and $v_0=255$ km/s (dashed) assuming $\delta=130$ keV. Comparing this to Fig.\ \ref{ratiovesc}, we see that it has the same features, but the effects are exaggerated.

The top and middle panel of Fig.\ \ref{v0fraction} shows the fractional change in $\sigma_n$ relative to the value at $v_0=220$ km/s for $v_0=195$ km/s (dotted) and $v_0=255$ km/s (dashed), assuming elastic and momentum dependent scattering. For clarity, in the top panel, we have only plotted the fractional change for CDMS II-Ge (blue) and XENON10 (red) since the fractional changes for CRESST-II and CDMS II-Si are similar to those of XENON10. In the middle panel, we have only plotted the change for CDMS II-Ge and CDMS II-Si.  In the top panel, for masses $>75$ GeV, the fractional change in $\sigma_n$ when increasing or decreasing $v_0$ remains at $\sim 10\%$ for all experiments. However, for lower masses, the fractional change at all experiments increases dramatically, reaching $\sim100\%$ at $10$ GeV. This sudden increase is caused by the horizontal shift in the exclusion limits which occurs for the same reasons as outlined in Section \ref{sectionvesc} above. In the middle panel, the CDMS II-Ge curves are similar to those in the top panel (although not shown, XENON10 and CRESST-II are also similar), but those for CDMS II-Si are quite different. This is because for momentum dependent scattering, the events at higher recoil energies are more important. To scatter with a larger recoil energy, a larger speed is required. For CDMS II-Si, this minimum speed is close to $v_{esc}$, so the limits are effected more when $v_0$ is varied, compared with elastic scattering.

We can use the formulas from Section \ref{sectionvesc} to estimate the vertical and horizontal shifts of the exclusion curves for inelastic scattering. In particular for XENON10 at $M_{\chi}=400$ GeV, $\delta=130$ keV and $E_R=35$ keV,\footref{first_footnote} we find $\Delta \sigma^{195km/s}_{220km/s}=1.59$ and $\Delta \sigma^{255km/s}_{220km/s}=-0.56$, in good agreement with the values $1.32$ and  $-0.60$ presented in the bottom panel of Fig. \ref{v0fraction}, while for the horizontal shift at $M_{\chi}=65$ GeV,  $\delta=130$ keV, $\Delta v_{\chi}=\pm35$ km/s and $v_{\chi}\sim700$ km/s we find $\Delta M_{\chi}\sim\mp5$ GeV. For XENON10 at $M_{\chi}=12$ GeV,  $\delta=0$ keV, $\Delta v_{\chi}=\pm35$ km/s and $v_{\chi}\sim500$ km/s we find $\Delta M_{\chi}\sim\mp1$ GeV. These estimates are all in reasonable agreement with what is observed in Fig.\ \ref{v0}.

\subsection{The local dark matter density}

We next consider the local dark matter density $\rho_{\chi}$. Unfortunately, this is the least well known astrophysical parameter with an often quoted uncertainty of a factor of $2$ or $3$ in the fiducial value $\rho_{\chi}=0.3$ GeV/cm$^3$ \cite{Gates:1995dw, Amsler:2008zzb}.  Two recent studies using different techniques found $\rho_{\chi}=0.3\pm0.1$ GeV/cm$^3$ \cite{Weber:2009pt} and $\rho_{\chi}=0.43\pm0.15$ GeV/cm$^3$ \cite{Salucci:2010qr}, consistent with $\rho_{\chi}=0.3$ GeV/cm$^3$ with an error of a factor of $2$. Dark matter N-body simulations from the Aquarius Project have shown that $\rho_{\chi}$ should be very smooth at the Sun's position, varying by less than $15\%$ at the $99.9\%$ confidence level from the average value over an ellipsoidal shell at the Sun's position \cite{Vogelsberger:2008qb}. Therefore we can be reasonably confident that the Earth is not sitting in a particularly over or under dense region of the halo.  

Since the combination $\rho_{\chi} \sigma_n$ enters into $dR/dE_R$ (Eqn.\ \ref{dRdE1}), the error in $\rho_{\chi}$ corresponds directly to an uncertainty of a factor of $2$ in the dark matter-neutron cross section $\sigma_n$, affecting all experiments equally. 

\subsection{Discussion: The correlation between $v_0$ and $\rho_{\chi}$}

\begin{figure*}[t]
\centering
\includegraphics[height=4.6in]{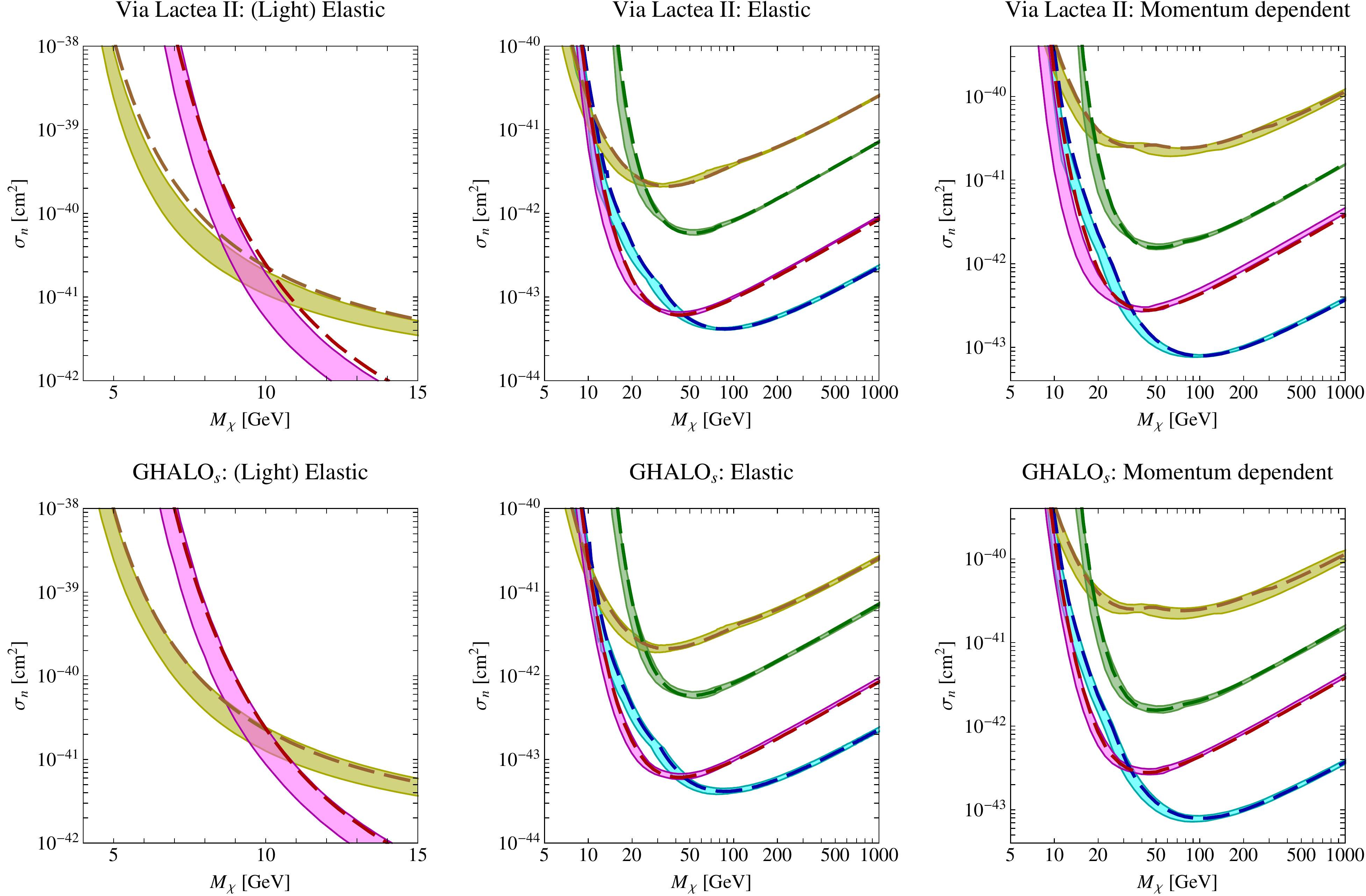}\caption{(Colour online). The exclusion limits from the Via Lactea II (upper panels) and GHALO$_{\text{s}}$ (lower panels) simulations. The legend is the same as Fig.\ \ref{v0}. The solid bands shows the range of limits from the velocity distribution of 100 random spheres centred at 8.5 kpc. For comparison, the dashed line show the limits from a Maxwell-Boltzmann velocity distribution with the same astrophysical parameters as found in the simulation halos. The Maxwell-Boltzmann distribution generally sets less constraining limits on $\sigma_n$.}
\label{simelastic}
\end{figure*}

We have followed the standard procedure of varying these astrophysical parameters independently, however, Refs.\ \cite{Reid:2004rd,McMillan:2009yr} found that there is a strong correlation between the Sun's circular speed $v_0$ and the distance from the galactic centre $R_0$: $v_0/R_0\approx29$ kms$^{-1}$kpc$^{-1}$. Since $\rho_{\chi}$ also depends on $R_0$, a more careful approach should take into account the change in $\rho_{\chi}$ as we vary $v_0$. Here we estimate this correction.

Assuming $v_0=230$ km/s and $\rho_{\chi}=0.3$ GeV/cm$^3$ at $R_0=8$ kpc, and assuming $\rho_{\chi}$ follows an NFW profile, when $\Delta R_0=\pm1$ kpc, we find $\Delta v_0\approx\pm30$ km/s and $\Delta \rho_{\chi}/\rho_{\chi}\approx\mp20\%$, similar to the range of $v_0$ explored in Section \ref{Suncirc}. Therefore for higher (lower) values of $v_0$, the exclusion curves in Fig.\ \ref{v0} are shifted upwards (downwards) by an additional $\sim20\%$. Comparing with Figs.\ \ref{v0} and \ref{v0fraction}, we see that this correction is of a similar size to what we found for $\Delta \sigma_n/\sigma_n$ when independently varying $v_0$ for the elastic and momentum dependent scattering case when $M_{\chi}\gtrsim30$ GeV, although the spread in the limits remains small. For light and inelastic dark matter, the change when varying $v_0$ independently is $\sim200\%$, therefore the spread in limits shown in Fig. \ref{v0} is overestimated by $\sim 10\%$, but still remains significant.


\section{Uncertainties in the form of the dark matter velocity distribution}\label{VLGHALO}

\begin{figure*}[t]
\centering
\includegraphics[height=3.3in]{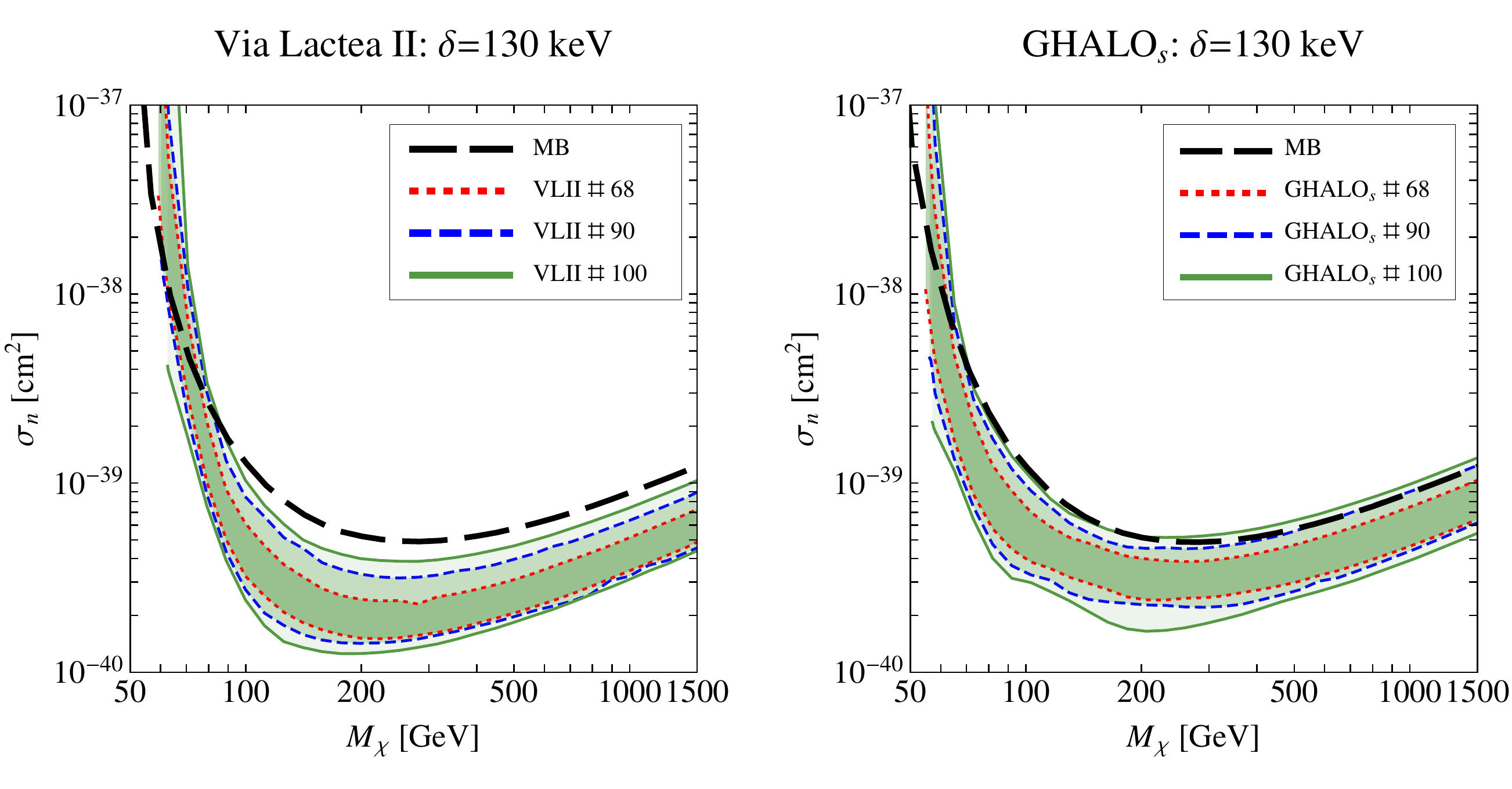}\caption{(Colour online). The exclusion limits for CRESST-II, assuming inelastic scattering from $100$ random spheres from the Via Lactea II (left panel) and GHALOs (right panel) simulations. The dotted red and dashed blue lines indicate the range of limits when we exclude five and sixteen of the highest and lowest cross sections at each mass. For comparison, the long-dashed black line shows the limits from a Maxwell-Boltzmann distribution with the same astrophysical parameters and dispersion. At $M_{\chi}\sim60$ GeV, we find that some of the spheres are not able to set any limits. As in Fig.\ \ref{simelastic}, we find that the Maxwell-Boltzmann distribution generally sets less constraining limits on $\sigma_n$.}
\label{siminelastic}
\end{figure*}

In the previous section we investigated the effect of astrophysical uncertainties whilst always assuming a Maxwell-Boltzmann distribution for the dark matter velocity distribution. Dark matter numerical N-body simulations have shown that a typical velocity distribution exhibits global (the overall shape) and local departures (`bumpy' features) from a Maxwell-Boltzmann distribution, and in this section we wish to investigate how these affect exclusion limits.

We use data extracted from the Via Lactea II (VLII) \cite{Diemand:2008in} and GHALO \cite{Stadel:2008pn} simulations available from \cite{Kuhlensite}. To facilitate a more direct comparison between the VLII and GHALO simulations, we use the GHALO$_{\text{s}}$ data from \cite{Kuhlen:2009vh} in which the maximum circular speed in the GHALO simulation has been rescaled to agree with that from VLII. As explained in \cite{Kuhlen:2009vh}, the lack of baryons in the simulations means that the dispersion of the extracted velocity distribution is lower than expected from a galaxy which contains baryons and dark matter, because it is expected that the baryons will create a deeper gravitational potential well, increasing the velocity dispersion.  Since we are interested in departures from a Maxwellian distribution, we will compare the simulation limits with those from a Maxwell-Boltzmann distribution with the same dispersion. This corresponds to using $v_0=184$ km/s in Eqn.\ \ref{fgal}. In the previous section we found that different astrophysical parameters can lead to significant deviations in the exclusion limits, therefore we will match these as closely as possible so that we are just observing the effects of the simulation distributions. Following \cite{Kuhlen:2009vh} we take $\rho_{\chi}=0.3$ GeV/cm$^3$, $\vec{v}_\odot=(10.0, 220+5.25 ,7.17)$ km/s and $v_{esc}=550$ km/s and $586$ km/s for VLII and GHALO$_{\text{s}}$ respectively.

With the simulation data, it is possible to examine the effects of local structure in the dark matter velocity distribution. This was done in \cite{Kuhlen:2009vh} by extracting the velocity distribution for one hundred spheres of radii $1.5$ kpc and $1$ kpc for VLII and GHALO$_{\text{s}}$ respectively, centred at a distance of $8.5$ kpc from the galactic centre. In Fig.\ \ref{simelastic} the thin bands show the exclusion curves from Via Lactea II (upper panels) and GHALO$_{\text{s}}$ (lower panels) assuming elastic and momentum dependent scattering for all one hundred spheres for CDMS II-Si (yellow), CDMS II-Ge (blue), CRESST-II (green) and XENON10 (red). We have checked the robustness of these bands by comparing the strongest and weakest limit with the sixteenth strongest and sixteenth weakest limit at each mass, and find that the limits are essentially the same. The dashed line shows the limits derived from a Maxwell-Boltzmann distribution with the astrophysical  and dispersion parameters above. We find that VLII and GHALO$_{\text{s}}$ produce similar limits, which are generally more stringent than those from a Maxwell-Boltzmann distribution. 

The inelastically scattering case has been studied in detail in \cite{Kuhlen:2009vh}, therefore in Fig.\ \ref{siminelastic} we only plot the limits for CRESST-II for $\delta=130$ keV (for CDMS II-Ge and XENON10 we find a similar spread in the limits). To give an estimate of what the typical limits are, we have plotted the limits excluding five and sixteen maximum and minimum cross sections at each mass as the blue dashed and red dotted bands respectively. As we might expect, the inelastic case shows much more of a spread in the cross section compared to the elastic case, and the limits from VLII and GHALO$_{\text{s}}$ show some differences. Interesting features to note are the bumpy features in the GHALO$_{\text{s}}$ exclusion limits, and near threshold we find that some spheres can not set any limits for both VLII and GHALO$_{\text{s}}$. The black dashed line shows the exclusion limits from a Maxwell-Boltzmann distribution which as in the elastic case, is typically more conservative than the limits from the simulations. 

We emphasise here that we have not been comparing the simulation limits with the SHM, in which the circular speed enters Eqn.\ \ref{fgal}, but rather to a best fit Maxwell-Boltzmann distribution which has the same velocity dispersion. Until simulations include baryonic effects, which will increase the dispersion, we urge caution in making direct comparisons to the SHM. The velocity distributions from the simulations are interesting since they show that the overall shape of the limits can change, rather than causing a spread about a central value as we found when varying the astrophysical parameters (cf Figs.\ \ref{v0} and \ref{siminelastic}). Furthermore since different experiments vary by different amounts, when comparing exclusion curves at one experiment with preferred regions of parameter space at another (for example, comparing CDMS II-Ge to the DAMA/LIBRA or CoGeNT preferred region), one should bear in mind that different velocity distributions may affect the limits from different experiments by differing amounts.
\section{Conclusions}

If we are to understand the nature of dark matter, it is vital that we have a full understanding of the astrophysical uncertainties affecting dark matter direct detection experiments. For spin-independent elastic and momentum dependent scattering with $M_{\chi}\gtrsim50$ GeV, we have shown that the exclusion limits are robust against variations in the galactic escape velocity $v_{esc}$ and the Sun's circular speed about the centre of the galaxy $v_0$ (right upper and left lower panels of Figs.\ \ref{v_esc} and \ref{v0}) and under realistic variations in the form of the velocity distribution (middle and right panels of Fig.\ \ref{simelastic}), with uncertainty $\sim 10\%$. The major uncertainty in this mass range arises from the error in the local dark matter density (a factor of $2$).

In comparison, for lighter masses, we found that uncertainties in $v_0$ and $v_{esc}$ and the velocity distributions from the numerical simulations can shift the exclusion curves horizontally by $\sim 1$ GeV at masses $M_{\chi}\sim10$ GeV (upper left panels of Figs.\ \ref{v_esc}, \ref{v0} and \ref{simelastic}). Similarly for inelastically scattering dark matter, we found the vertical shift in the exclusion curves when varying $v_0$ or $v_{esc}$ is large ($\sim100\%$), with lighter target experiments such as CDMS II being particularly affected. Variations in the velocity distribution can also lead to significant changes, as we explicitly demonstrated in Fig.\ \ref{siminelastic} for the CRESST-II experiment.

Inelastic and light dark matter are particularly sensitive to astrophysical uncertainties because the minimum speed needed to scatter is just below the galactic escape speed, so experiments only sample the tail of the velocity distribution. Even though the recoil spectrum for momentum dependent dark matter is different from elastically scattering dark matter (Fig.\ \ref{3dRdE}) they both respond to astrophysical uncertainties in a similar fashion for germanium, xenon and tungsten targets, because $v_{min}$ remains far from $v_{esc}$. Therefore models which only sample the tail of the velocity distribution should carefully examine the effect of astrophysical uncertainties on their limits.

\begin{acknowledgments}
It is a pleasure to thank James Binney, John Magorrian and Subir Sarkar for discussions on the astrophysical uncertainties, Jody Cooley, Kaixuan Ni and Peter Sorenson for clarifying experimental details, Micheal Kuhlen, Mark Vogelsberger and Neal Weiner for discussions on the dark matter N-body simulations and Sebastian Cassel, Mads Frandsen, John March-Russell and Matthew McCullough for  many fruitful discussions. CM is supported by a STFC postgraduate studentship.
\end{acknowledgments}

\appendix

\section{Experimental Details}\label{experiments}

In this Appendix we discuss the four experiments which we will use as our reference experiments, and highlight particular uncertainties associated with each. These experiments all publish unbinned data, so we use Yellin's `pmax' method \cite{yellin} to calculate the limit on the dark matter-nucleon cross section, since it generally provides a stronger constraint than unbinned methods for the case of an unknown background. It produces similar limits to the `optimum interval' method used by CDMS II, the `maximum gap' method used by CRESST-II and XENON10 (in their elastic dark matter analysis), and was later used by XENON10 (in their inelastic dark matter reanalysis).

\subsection{CDMS II-Ge}

We initially consider the four published running periods (runs 118, 119, 123-124 and 125-128) by the CDMS II collaboration for scattering off germanium in the energy range $10-100$ keV \cite{Akerib:2004fq, Akerib:2005kh, Ahmed:2008eu, Ahmed:2009zw}. The first analysis saw one event at $10.5$ keV; the second, one event at $64$ keV; the third, no events; and the fourth, notably saw two events at $12.3$ keV and $15.5$ keV. 

It was realised that there was an error in the published analysis of the first two runs (118 and 119) and a combined reanalysis was performed, which had redesigned cuts, and lowered the energy threshold from $10$ keV to $5$ keV \cite{Ogburn:2008zz}. After the cuts were applied, it was found that the two events at $10.5$ keV and $64$ keV were no longer present, but two new events at $5.3$ keV and $7.3$ keV passed all the cuts. These were not thought to be real dark matter events. The $5.3$ keV event is very close to the energy threshold so was thought to be from background leakage, while the $7.3$ keV event had an unphysical negative phonon delay, but was passed because of a pathology in the cut definitions. We follow CDMS by using this reanalysis when setting limits rather than the two published runs.\footnote{We have checked the exclusion limits without the unphysical event at $7.3$ keV and the differences are small.} It should be noted that due to the lower energy threshold of these first two runs ($5$ keV rather than $10$ keV), CDMS II-Ge is able to place stronger limits on light dark matter than is often calculated. Similarly, the removal of the event at $64$ keV leads to stronger limits being set on inelastic dark matter since this event lay near the peak of the recoil spectrum $dR/dE_R$ \cite{Chang:2008gd}.

In the analysis of the final run, an improved measurement of the germanium detector mass found a $9\%$ decrease in the effective exposure. In addition, a correction of the neutron systematic errors led to a $4.5\%$ increase in the effective exposure \cite{JCooley}. We therefore multiply the published exposures (excluding the final run analysis) by $0.955$.

The effective germanium exposure of the combined analysis of runs 118 and 119 is given in Figure 10.1 of \cite{Ogburn:2008zz}. We find a good fit to this curve for $E_R\geq5$ keV is given by:
\begin{equation*}
\frac{Exp}{0.955}=80.1-0.31 E_R-225.52 e^{-E_R/2.09}-74.46 e^{-E_R/20.0}
\end{equation*}

For the exposure of the 2008 analysis (runs 123-124), we use $0.995\times397.8$ kg-days and use the fit to the efficiency from \cite{Savage:2008er}.

For the 2009 analysis (runs 125-128), we use an exposure of $612$ kg-days and find
\begin{equation*}
eff=0.38-1.3\times10^{-3}E_R-0.17 e^{-E_R/7.98}-2.96 e^{-Er/2.86}
\end{equation*}
is a good fit to the efficiency.\footnote{Figure can be found in the online supplementary material of \cite{Ahmed:2009zw}.}

\subsection{CDMS II-Si}

The silicon data from the first two runs was also reanalysed and two events at $34.9$ keV and $94.1$ keV passed all the cuts \cite{Ogburn:2008zz}. The 2008 analysis saw no events \cite{Filippini:2008zz}. These runs had a low energy threshold of $5$ keV and $7$ keV respectively, so again, are capable of setting strong limits at low masses.

The effective silicon exposure of the combined analysis of runs 118 and 119 is given in Figure 10.1 of \cite{Ogburn:2008zz}. We find a good fit to this curve for $5\text{ keV}\leq E_R\leq20$ keV is given by:
\begin{equation*}
Exp_{5-20}=11.29-28.7e^{-E_R/5.34}
\end{equation*}
and for $20\text{ keV}< E_R\leq100$ keV is given by:
\begin{equation*}
Exp_{20-100}=(0.55-0.63e^{-E_R/15.1})(0.061E_R+29.5)
\end{equation*}

For the exposure of the 2008 analysis, we use $53.47$ kg-days and fit to the efficiency from Fig. 5.13 of \cite{Bailey}.
\begin{equation*}
eff = \left\{
\begin{array}{l@{\quad}l}
0 & E_R < 7\text{ keV} \\
2 E_R/30-7/15 & 7\text{ keV}<E_R < 10\text{ keV} \\
0.0075 E_R+0.125 & 10\text{ keV}<E_R < 14\text{ keV} \\
0.09 E_R-1.03 & 14\text{ keV}<E_R < 15\text{ keV} \\
0.02 E_R+0.02 & 15\text{ keV}<E_R < 19\text{ keV} \\
0.4 & 19\text{ keV}<E_R < 100\text{ keV} \\
\end{array}\right.
\end{equation*}

\subsection{CRESST-II}

CRESST-II have released data from two modules from the prototype phase \cite{Angloher:2004tr} and two modules from the commissioning phase \cite{Angloher:2008jj}. Here we only include the seven events in the energy range $10-100$ keV from the commissioning phase, since there was a significant upgrade to the experiment after the prototype phase. Combining the data leads to significantly weaker limits \cite{SchmidtHoberg:2009gn}.

\subsection{XENON10}

Finally we consider the liquid xenon based experiment XENON10, which combines scintillation and ionisation measurements to infer the nucleus recoil energy $E_R$. There have been two analyses performed; the first considered only elastically scattering dark matter \cite{Angle:2007uj} and saw 10 events in the energy range $4.5-26.9$ keV; the second analysis, which used the same data but with redesigned cuts considered inelastically scattering dark matter \cite{Angle:2009xb} and saw 13 events in an energy range  extended to $75$ keV. We use an energy resolution for XENON10 given in \cite{Savage:2008er}.

The quantity $\mathcal{L}_{eff}$ is needed to convert from the measured scintillation energy (measured in keVee), to the nuclear recoil energy (measured in keV). Unfortunately $\mathcal{L}_{eff}$ is poorly determined and presents the largest systematic uncertainty in the results of the XENON10 experiment. The original analysis, and the energies quoted above assume a constant value $\mathcal{L}_{eff}=0.19$, however more recently, \cite{Aprile:2008rc, Sorensen:2008ec,Manzur:2009hp} have determined that $\mathcal{L}_{eff}$ decreases at low energies, but there is still some uncertainty in its value. Refs. \cite{Aprile:2008rc, Sorensen:2008ec} found $\mathcal{L}_{eff}\sim0.15$ for $E_R<10$ keV, while Ref. \cite{Manzur:2009hp} found that it continues to decrease from $\mathcal{L}_{eff}\sim0.15$ at $E_R=10$ keV to $\mathcal{L}_{eff}\sim0.10$ at $E_R=5$ keV. At $M_{\chi}\sim10$ GeV, the exclusion limits from $\mathcal{L}_{eff}=0.19$ and $\mathcal{L}_{eff}$ from Manzur et al.\ differ by more than an order of magnitude \cite{Manzur:2009hp}. 

When setting limits in this paper, we use the average value of $\mathcal{L}_{eff}$ from  \cite{Aprile:2008rc, Sorensen:2008ec,Manzur:2009hp} and use the elastic analysis to set exclusion limits for momentum dependent and elastically scattering dark matter, and the inelastic analysis when setting limits for inelastically scattering dark matter.

We end this section by mentioning another liquid xenon based experiment,  ZEPLIN-III \cite{Lebedenko:2008gb}, which uses a value of $\mathcal{L}_{eff}$ which is smaller at low recoil energy than the three measurements mentioned above. Therefore in contrast to XENON10, we would expect the ZEPLIN-III limits to be stronger at low masses if they were to use $\mathcal{L}_{eff}$ from Manzur et al.\  However we don't reproduce their limits here since they take into account background subtraction which is beyond our capabilities.

\section{Analytic formula for $\zeta(E_R)$}\label{analytic_integration}
Here we present an analytic formula for $\zeta(E_R)$, defined by
\begin{equation}\label{zetaER2}
\zeta(E_R)=\int^\infty_{v_{min}} \frac{d^3v}{v} f(\vec{v}+\vec{v}_e),
\end{equation}
where
\begin{equation}\label{fgal2}
f(\vec{v}) = \left\{
\begin{array}{l@{\qquad}l}
\frac{1}{N} \left(e^{-v^2 / v_0^2} - \beta e^{-v_{esc}^2 / v_0^2}\right)
& v < v_{esc} \\
0 & v > v_{esc}
\end{array}\right.
\end{equation}
and $\beta=0$ or $1$, depending on whether the exponential cutoff is desired. An analytic formula for $\zeta(E_R)$ with $\beta=0$ was presented in \cite{Savage:2006qr} and we agree with their results in this limit. In this paper, we take $\beta=1$.

It is convenient to define $x_{esc}=v_{esc}/v_0$,  $x_{min}=v_{min}/v_0$ and  $x_e=v_e/v_0$, where $v_i=\lvert \vec{v}_i\rvert$.
\begin{widetext}
The normalisation constant $N$ is given by:
\begin{equation}
N=\pi^{3/2} v_0^3 \left[\operatorname{erf}(x_{esc}) - \frac{4}{\sqrt{\pi}}e^{-x_{esc}^2}\left(\frac{x_{esc}}{2}+\frac{\beta x_{esc}^3}{3} \right) \right]
\end{equation}
If $x_e+x_{min}<x_{esc}$:
\begin{equation}
\zeta(E_R)=\frac{\pi^{3/2} v_0^2}{2 N x_e} \left[\operatorname{erf}(x_{min}+x_{e})-\operatorname{erf}(x_{min}-x_{e}) 
 - \frac{4 x_e}{\sqrt{\pi}} e^{-x_{esc}^2}\left(1+\beta (x_{esc}^2-\frac{x_{e}^2}{3}-x_{min}^2)   \right)
\right]
\end{equation}
If $x_{min}>\lvert x_{esc}-x_e\rvert$ and $x_e+x_{esc}>x_{min}$:
\begin{align}\label{analytic2}
\zeta(E_R)&=\frac{\pi^{3/2} v_0^2 }{2 N x_e} \Bigg[ \operatorname{erf}(x_{esc})+\operatorname{erf}(x_e-x_{min})  \notag \\
& - \frac{2}{\sqrt{\pi}} e^{-x_{esc}^2}\left(x_{esc}+x_e-x_{min}-\frac{\beta}{3}(x_e-2 x_{esc}-x_{min})(x_{esc}+x_{e}-x_{min})^2 \right) \Bigg]
\end{align}
If $x_e>x_{min}+x_{esc}$: 
\begin{equation}
\zeta(E_R)=\frac{1}{v_0 x_e}
\end{equation}
If $x_e+x_{esc}<x_{min}$:
\begin{equation}\label{analytic4}
\zeta(E_R)=0
\end{equation}
\end{widetext}

\bibliography{myrefs}

\end{document}